\documentclass[twocolumn,prd,superscriptaddress,showpacs,floatfix
,preprintnumbers,nofootinbib]{revtex4}

\usepackage{epsfig}
\usepackage{bm}
\usepackage{color}

\begin{document}

\title{Reconstructing the supernova bounce time with neutrinos
in IceCube}

\author{Francis Halzen}
\affiliation{Department of Physics, University of Wisconsin,
Madison, WI 53706, USA}

\author{Georg~G.~Raffelt}
\affiliation{Max-Planck-Institut f\"ur Physik
(Werner-Heisenberg-Institut), F\"ohringer Ring 6, 80805 M\"unchen,
Germany}

\date{17 August 2009, finalized 26 September 2009}

\preprint{MPP-2009-152}

\begin{abstract}
Generic model predictions for the early neutrino signal of a
core-collapse supernova (SN) imply that IceCube can reconstruct the
bounce to within about $\pm 3.5$~ms at 95\% CL (assumed SN distance
10~kpc), relevant for coincidence with gravitational-wave detectors.
The timing uncertainty scales approximately with distance-squared.
The offset between true and reconstructed bounce time of up to several
ms depends on the neutrino flavor oscillation scenario. Our work
extends the recent study of Pagliaroli
et~al.\ [Phys.\ Rev.\ Lett.\ 103, 031102 (2009)] and demonstrates
IceCube's superb timing capabilities for neutrinos from the next
nearby~SN.
\end{abstract}

\maketitle

\section{Introduction}                        \label{sec:introduction}

The high-statistics neutrino observation from the next nearby
supernova (SN) will provide a bonanza of information about the
astrophysics of core-collapse phenomena and neutrino properties. In a
recent Physical Review Letter a strong case was made for the
importance of coincidence measurements between gravitational
wave and neutrino signals from SN core
bounce~\cite{Pagliaroli:2009qy}. The largest existing SN neutrino
detectors are Super-Kamiokande and IceCube, reaching to a distance of
about 100~kpc. The expected distribution of galactic SNe
drops off quickly beyond about 20~kpc \cite{Mirizzi:2006xx}. At this
``pessimistic'' distance, Super-Kamiokande can time the bounce to
within a few tens of milliseconds, an interval comparable to the
expected duration of the gravitational wave
burst~\cite{Pagliaroli:2009qy}.

We here extend this study to IceCube, a high-statistics SN neutrino
detector that would have seen the SN~1987A neutrino signal with
$5\,\sigma$ significance. For galactic SNe, the large rate of
uncorrelated Cherenkov photons provides excellent time-structure
information.

Following Ref.~\cite{Pagliaroli:2009qy} we note that neutrino masses
are small enough to neglect time-of-flight effects. Recalling that
the diameter of the Earth is 42~ms, millisecond-coincidence
measurements between detectors at different geographic locations
requires determining the SN direction either by astronomical
observations or by the electron-scattering signal in
Super-Kamiokande.

\section{Supernova neutrinos in IceCube}           \label{sec:IceCube}

When SN neutrinos stream through water or ice, Cherenkov light is
generated, primarily by the secondary positrons from inverse beta
decay $\bar\nu_e+p\to n+e^+$. While fine-grained detectors
reconstruct individual neutrinos on an event-by-event basis, IceCube
only picks up the average Cherenkov glow of the ice. To estimate the
detection rate we follow Ref.~\cite{Dighe:2003be}, augmented with
the latest IceCube efficiencies~\cite{Kowarik:2009qr}. The complete
detector will have 4800 optical modules (OMs) and the data are read
out in 1.6384~ms bins, implying a total event rate~of
\begin{equation}
R_{\bar\nu_e}=186~{\rm bin}^{\!-1}\,
L_{52}\,D_{10}^{-2}\,
\langle E_{15}^3\rangle/\langle E_{15}\rangle\,,
\end{equation}
where $L_{52}=L_{\bar\nu_e}/10^{52}~{\rm erg}~{\rm s}^{-1}$,
$D_{10}=D/10~{\rm kpc}$ and $E_{15}=E_{\bar\nu_e}/15~{\rm MeV}$.

The peak luminosity reaches $L_{52}=2$--5 and at that time $\langle
E_{\bar\nu_e}\rangle\approx15$~MeV. For a thermal or slightly
pinched spectrum, $\langle E_{\bar\nu_e}^3\rangle/\langle
E_{\bar\nu_e}\rangle^3\approx2$. Altogether, we expect
\begin{equation}
R_{\bar\nu_e}^{\rm max}=1.5\times10^3~{\rm bin}^{\!-1}
\end{equation}
as a typical peak event rate for a SN at 10~kpc.

This signal is to be compared with a background of 280~s$^{-1}$ in
each OM~\cite{Kowarik:2009qr}, corresponding for 4800~OMs to
\begin{equation}
R_{0}=2.20\times10^{3}~{\rm bin}^{\!-1}
\end{equation}
with an rms fluctuation of $47~{\rm bin}^{\!-1}$. Therefore, the
SN-induced ``correlated noise'' in the entire detector is highly
significant~\cite{Pryor:1987tz, Halzen:1995ex, Dighe:2003be}. From
any one neutrino interacting in the ice, at most one Cherenkov photon
is picked up: the signal counts are entirely uncorrelated.

The prompt $\nu_e$ burst immediately after bounce produces a peak
rate of about $100~{\rm bin}^{\!-1}$, lasting for several ms. Its
impact on the early count rate depends sensitively on the flavor
oscillation scenario (see below).

\section{Early Neutrino Emission}         \label{sec:NeutrinoEmission}

SN models suggest that neutrino emission for the first 20~ms after
core bounce depends little on model assumptions or input
physics~\cite{Kachelriess:2004ds}, although beyond this early phase
the accretion rate and therefore neutrino emission depends strongly,
for example, on the progenitor mass profile. The prompt $\nu_e$ burst
reaches its peak at 5--7~ms post bounce, which also marks the onset of
$\bar\nu_e$ emission that is initially suppressed by the large $\nu_e$
chemical potential. Up to about 20~ms post bounce $L_{\bar\nu_e}$
rises roughly linearly. We thus represent the early IceCube signal
as~\cite{Pagliaroli:2009qy}
\begin{equation}\label{eq:risemodel}
R_{\bar\nu_e}=R_{\bar\nu_e}^{\rm max}
\times\cases{0&for $t<t_{\rm r}$\cr
1-e^{-(t-t_{\rm r})/\tau_{\rm r}}&for $t>t_{\rm r}$\cr}
\end{equation}
with $t_{\rm r}=6$~ms, $\tau_{\rm r}=50$~ms and $R_{\bar\nu_e}^{\rm
max}=1.5\times10^3~{\rm bin}^{\!-1}$. These parameters also provide
an excellent fit to the first 100~ms of a numerical model from the
Garching group~\cite{Marek:2008qi} that is available to us.

We may compare these assumptions with the early-phase models of
Ref.~\cite{Kachelriess:2004ds}. $L_{\bar\nu_e}$ rises nearly
linearly to $L_{52}=1.5$--2 within 10~ms. The evolution of $\langle
E_{\bar\nu_e}\rangle_{\rm RMS}=(\langle
E_{\bar\nu_e}^3\rangle/\langle E_{\bar\nu_e}\rangle)^{1/2}$ is also
shown, a common quantity in SN physics that characterizes, for
example, the efficiency of energy deposition; the IceCube rate is
proportional to $\langle E_{\bar\nu_e}\rangle_{\rm RMS}^2$. At 10~ms
after onset, $\langle E_{\bar\nu_e}\rangle_{\rm RMS}$ reaches
15~MeV, implying $\langle E_{15}^3\rangle/\langle E_{15}\rangle=1$.
We thus estimate 10~ms after onset a rate of 280--$370~{\rm
bin}^{\!-1}$, to be compared with $270~{\rm bin}^{\!-1}$ from
Eq.~(\ref{eq:risemodel}). Therefore, our assumed signal rise is on
the conservative side.

Of course, the early models do not fix $\tau_{\rm r}$ and
$R_{\bar\nu_e}^{\rm max}$ separately; the crucial parameters are
$t_{\rm r}$ and $R_{\bar\nu_e}^{\rm max}/\tau_{\rm r}$. The maximum
rate that is reached long after bounce is not relevant for
determining the onset of the signal.

If flavor oscillations swap the $\bar\nu_e$ flux with $\bar\nu_x$
(some combination of $\bar\nu_\mu$ and $\bar\nu_\tau$), the rise
begins earlier because the large $\nu_e$ chemical potential during
the prompt $\nu_e$ burst does not suppress the early emission of
$\bar\nu_x$ \cite{Kachelriess:2004ds}. Moreover, the rise time is
faster, $\langle E\rangle_{\rm RMS}$ larger, and the maximum
luminosity smaller. We use Eq.~(\ref{eq:risemodel}) also for
$R_{\bar\nu_x}$ with $t_{\rm r}=0$, $\tau_{\rm r}=25$~ms, and
$R_{\bar\nu_x}^{\rm max}=1.0\times10^3~{\rm bin}^{\!-1}$.

Flavor oscillations are unavoidable and have been studied, for early
neutrino emission, in Ref.~\cite{Kachelriess:2004ds}. Assuming the
normal mass hierarchy, $\sin^2\Theta_{13}\agt10^{-3}$, no collective
oscillations,\footnote{In the normal hierarchy, collective
oscillation effects are usually absent. It has not been studied,
however, if the early neutrino signal can produce multiple splits
that can arise also in the normal hierarchy~\cite{Dasgupta:2009mg}.
Moreover, for a low-mass progenitor collective phenomena can be
important if the MSW resonances occur close to the neutrino
sphere~\cite{Duan:2007sh, Dasgupta:2008cd}.} and a direct
observation without Earth effects, Table~I of
Ref.~\cite{Kachelriess:2004ds} reveals that the $\nu_e$ burst would
be completely swapped and thus nearly invisible because the $\nu_x
e^-$ elastic scattering cross section is much smaller than that of
$\nu_e$. The survival probability of $\bar\nu_e$ would be
$\cos^2\Theta_{12}\approx2/3$ with $\Theta_{12}$ the ``solar''
mixing angle. Therefore, the effective detection rate would be
$\frac{2}{3}\,R_{\bar\nu_e}+\frac{1}{3}\,R_{\bar\nu_x}$. We use this
case as our main example.

\section{Reconstructing the Signal Onset}   \label{sec:reconstruction}

\begin{figure}
\includegraphics[width=1.0\columnwidth]{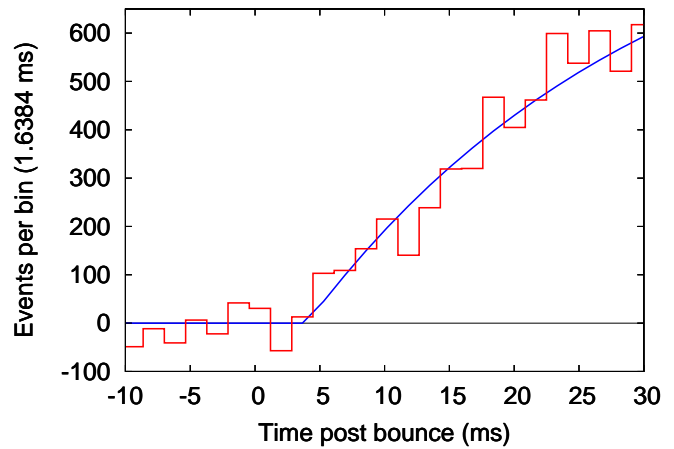}
\caption{Typical Monte Carlo realization (red histogram)
and reconstructed fit (blue line) for the benchmark case
discussed in the text for a SN at 10~kpc.\label{fig:simulation}}
\end{figure}

A typical Monte Carlo realization of the IceCube signal for our
example is shown in Fig.~\ref{fig:simulation}. One can determine the
signal onset $t_0$ within a few ms by naked eye. For a SN closer
than our standard distance of 10~kpc, one can follow details of the
neutrino light curve without any fit.

One can not separate the $\bar\nu_e$ and $\bar\nu_x$ components for
the example of Fig.~\ref{fig:simulation}. Therefore, we reconstruct a
fit with a single component of the form Eq.~(\ref{eq:risemodel}),
assuming the zero-signal background is well known and not fitted
here. Using a time interval until 100~ms post bounce, we reconstruct
$t_0=3.2\pm1.0$~ms ($1\sigma$). If we use only data until 33~ms post
bounce we find $t_0=3.0\pm1.7$~ms.  Indeed, if one fits
Eq.~(\ref{eq:risemodel}) on an interval that ends long before the
plateau is reached, we effectively fit a second order polynomial with
a positive slope and negative second derivative at~$t_{\rm r}$,
whereas the plateau itself is poorly fitted and its assumed value
plays little role.  Depending on the distance of the SN one will fit
more or fewer details of the overall neutrino light curve and there
may be more efficient estimators for $t_{\rm r}$.  Our example only
provides a rough impression of what IceCube can do.

The reconstruction uncertainty of $t_0$ scales approximately with
neutrino flux, i.e., with SN distance squared. The number of excess
events above background marking the onset of the signal has to be
compared with the background fluctuations. Therefore, a significant
number of excess events above background requires a longer
integration period if the flux is smaller, explaining this scaling
behavior.

The interpretation of $t_0$ relative to the true bounce time depends
on the flavor oscillation scenario realized in nature. This is
influenced by many factors: The value of $\Theta_{13}$, the mass
ordering, the role of collective oscillation effects, and the
distance traveled in the Earth. Combining the signal from different
detectors, using future laboratory information on neutrino
parameters, and perhaps the very coincidence with a
gravitational-wave signal may allow one to disentangle some of these
features. However, as a first rough estimate it is sufficient to say
that the reconstructed $t_0$ tends to be systematically delayed
relative to the bounce time by no more than a few ms. The
statistical uncertainty of the $t_0$ reconstruction does not depend
strongly on the oscillation scenario.

\section{Discussion}                   \label{sec:summary}

The authors of Ref.~\cite{Pagliaroli:2009qy} used a SN emission
model based on a two-component fit of the sparse SN~1987A data and
described the overall neutrino signal in terms of several parameters
which they say, after Eq.~(5), are at odds with theoretical
expectations. However, their parameter $T_a=2.4$~MeV, describing the
temperature of the neutrino-emitting gas during the accretion phase,
gives $\langle E_{\bar\nu_e}\rangle=5.2\,T_a=12.5$~MeV (see
paragraph after Eq.~15 in the published version of
Ref.~\cite{Pagliaroli:2008ur}) and thus is virtually identical to
the corresponding $\langle E_{\bar\nu_e}\rangle_{\rm RMS}$ from
Ref.~\cite{Kachelriess:2004ds} at 20~ms post bounce. In other words,
while the SN~1987A implied $\bar\nu_e$ energies of
Ref.~\cite{Pagliaroli:2009qy} are lower than theoretical
expectations for the overall accretion phase, in the absence of
flavor oscillations they agree nicely with the models of
Ref.~\cite{Kachelriess:2004ds} for the crucial first 20~ms. However,
the chosen rise time $\tau_{\rm r}=100$~ms is very long compared
with the early-time models of Ref.~\cite{Kachelriess:2004ds}.

In Ref.~\cite{Kachelriess:2004ds} the early neutrino signal after
bounce was systematically studied for different input assumptions
(progenitor mass, equation of state, neutrino opacities), leading to
very similar results. On the other hand, one finds significantly
different numerical examples in the literature. In
Ref.~\cite{Mezzacappa:2000jb} the $\bar\nu_e$ luminosity rises to
$2\times10^{52}~{\rm erg~s}^{-1}$ after as much as 50~ms. In
Ref.~\cite{Fryer:2006uv}, a peak value of only $1\times10^{52}~{\rm
erg~s}^{-1}$ is reached and the rise within 10--20~ms after bounce
is small. In Ref.~\cite{Herant:1994dd} the $\bar\nu_e$ signal begins
rising as late as 12~ms after the maximum of the $\nu_e$ burst.
There are many differences between these and other models in terms
of numerical approach and input physics. It would be extremely
useful if another group would investigate the early neutrino signal
in the spirit of Ref.~\cite{Kachelriess:2004ds} for a range of
physical assumptions and with attention to numerical details that
may influence the early-time behavior.

In view of the large range of possible distances to the next nearby
SN and concomitant flux differences, these uncertainties do not
change our overall conclusions. In our fiducial example IceCube can
reconstruct the signal onset within $\pm 6$--7~ms at $1\sigma$ CL
for a SN at 20~kpc, comparable to what Ref.~\cite{Pagliaroli:2009qy}
found for Super-Kamiokande. Ideally, of course, one would combine
the measurements from several detectors.

A gravitational wave measurement of the core bounce in coincidence
with neutrino onset would be of obvious astrophysical importance. In
addition one could test the weak equivalence principle. Both
neutrinos and gravitational waves should suffer the same Shapiro
time delay in the gravitational potential of the galaxy. For
SN~1987A in the Large Magellanic Cloud, this delay was a few months.
The coincidence of the neutrino burst with the rise of the light
curve within a few hours proved an equal Shapiro delay for photons
and neutrinos to within about $10^{-3}$ \cite{Longo:1987gc,
Krauss:1987me}. A millisecond-scale coincidence between neutrinos
and gravitational waves would extend and refine this test, in detail
depending on the location of the SN.

If the SN can not be located by astronomical means because of
obscuration, the electron-recoil signal in Super-Kamiokande or a
future megatonne water Cherenkov detector is the method of
choice~\cite{Beacom:1998fj, Tomas:2003xn}, whereas arrival-time
triangulation was dismissed. With IceCube almost complete, the
situation has changed and triangulation can play a useful role after
all, at least for a not-too-distant SN. Moreover, in Europe a big
detector of the LAGUNA-class \cite{Autiero:2007zj} may become
available (representing one of three possible large-scale detector
types) and a megatonne water-Cherenkov detector may be built in North
America.  Arrival-time triangulation between large-scale neutrino
detectors on different continents could become viable.

It would be a worthwhile exercise to study the possibilities of
bounce timing, based on common astrophysical assumptions, for
different detector types and what can be learnt from a combined
analysis. To this end it would be worthwhile if groups other than
the Garching one would systematically study the numerical early-time
neutrino signal to judge if indeed the dependence on physical
assumptions is as small as found in Ref.~\cite{Kachelriess:2004ds}.

\begin{acknowledgments}
We thank B.~Dasgupta, H.-T.~Janka and A.~Mirizzi for discussions.
This research was supported in part by the U.S.\ National Science
Foundation under Grants No.\ OPP-0236449 and PHY-0354776 and by the
Alexander von Humboldt Foundation in Germany (F.H.) and
by the Deutsche Forschungsgemeinschaft under grant
TR-27 ``Neutrinos and Beyond'' and the Cluster of Excellence
``Origin and Structure of the Universe'' (G.R.).
\end{acknowledgments}


\end{document}